\documentclass[runningheads]{llncs}

\usepackage{graphicx}
\usepackage{times}  %Required
\usepackage{helvet}  %Required
\usepackage{courier}  %Required
\usepackage{url}  %Required

\usepackage{booktabs} % For formal tables
\usepackage{algorithm,algorithmic,multirow,epstopdf,subfigure}
\usepackage{amsfonts}

\usepackage{multirow}
\usepackage{amsthm}

\theoremstyle{plain}

\newcommand{\methodfull}{RNE}

\begin{document}
\title{ RNE: A Scalable Network Embedding for Billion-scale Recommendation }
%
%\titlerunning{Abbreviated paper title}
% If the paper title is too long for the running head, you can set
% an abbreviated paper title here
%
\author{Jianbin Lin$^1$, Daixin Wang$^{1,2}$, Lu Guan$^3$, Yin Zhao$^3$, Binqiang Zhao$^3$, Jun Zhou$^1$, Xiaolong Li$^1$, Yuan Qi$^1$}
%
%\authorrunning{F. Author et al.}
% First names are abbreviated in the running head.
% If there are more than two authors, 'et al.' is used.
%
\institute{Ant Financial Services Group, Hangzhou, China \and
           Computer Science and Technology, Tsinghua University, Beijing, China \and
           Alibaba Group, Hangzhou, China}
%Princeton University, Princeton NJ 08544, USA \and
%Springer Heidelberg, Tiergartenstr. 17, 69121 Heidelberg, Germany
%\email{lncs@springer.com}\\
%\url{http://www.springer.com/gp/computer-science/lncs} \and
%ABC Institute, Rupert-Karls-University Heidelberg, Heidelberg, Germany\\
%\email{\{abc,lncs\}@uni-heidelberg.de}}
%
\maketitle              % typeset the header of the contribution
\begin{abstract}
Nowadays designing a real recommendation system has been a critical problem for both academic and industry. However, due to the huge number of users and items, the diversity and dynamic property of the user interest, how to design a scalable recommendation system, which is able to efficiently produce effective and diverse recommendation results on billion-scale scenarios, is still a challenging and open problem for existing methods. In this paper, given the user-item interaction graph, we propose \methodfull, a data-efficient \underline{R}ecommendation-based \underline{N}etwork \underline{E}mbedding method, to give personalized and diverse items to users. Specifically, we propose a diversity- and dynamics-aware neighbor sampling method for network embedding. On the one hand, the method is able to preserve the local structure between the users and items while modeling the diversity and dynamic property of the user interest to boost the recommendation quality. On the other hand the sampling method can reduce the complexity of the whole method theoretically to make it possible for billion-scale recommendation. We also implement the designed algorithm in a distributed way to further improves its scalability. Experimentally, we deploy \methodfull{} on a recommendation scenario of Taobao, the largest E-commerce platform in China, and train it on a billion-scale user-item graph. As is shown on several online metrics on A/B testing, \methodfull{} is able to achieve both high-quality and diverse results compared with CF-based methods. We also conduct the offline experiments on Pinterest dataset comparing with several state-of-the-art recommendation methods and network embedding methods. The results demonstrate that our method is able to produce a good result while runs much faster than the baseline methods. 
\end{abstract}

\vspace{-20pt}
\section{Introduction}
\label{sec:introduction}
\vspace{-10pt}
With the exponential growth of data and information on the Internet, recommendation system plays a critical role in reducing information overload. Recommendation systems are widely deployed on many online services, including E-commerce, social networks and online news systems. How to design an effective recommendation system has been a fundamental problem in both academia and industry.

The key for recommendation system is to model the users' preferences based on their interactions (e.g., clicks and rating) with the items. One of the most popular recommendation methods are known as collaborative filtering (CF) \cite{Herlocker2004Evaluating}. Its basic idea is to match the users with similar item preferences. Among the various collaborative filtering methods, matrix factorization \cite{Harvey2011Bayesian,zhou2017enhancing} is the mostly used one. However, these matrix factorization based methods are regarded as the linear methods, which are difficult to model the user-item interactions. Then following works use the deep neural networks to model the user-item relationships \cite{He2017Neural}. Despite of their success, these CF-based methods only aim to model the direct links, i.e. the first-order relationship between the users and items. However, for a graph, only preserving the first-order relationships between the nodes is not enough to characterize the network structure and thus cannot achieve good performance \cite{cui2018survey,Cao2015GraRep}.

To preserve the second-order local structure in the networks, network embedding is an effective way \cite{cui2018survey}. Network embedding aims to embed nodes into a low-dimensional vector space with the goal of capturing the low-order and high-order topological characteristics in graphs \cite{perozzi2014deepwalk,tang2015line,Grover2016node2vec,Wang2016Structural,zhang2018billion}. Although network embedding is able to incorporate local structures, they mainly target on tasks of common link prediction and node classification. Few of them deal with the task of recommendation and thus they seldom consider some specific properties of recommendation, which makes them difficult to get a good performance on recommendation. Last but not least, few of these methods can be applied to the billion-scale networks. 

To extend network embedding to recommendation, we meet three challenges. (1) Diversity of user interest. User interest is always diverse and the diverse recommendation can help users explore new items of interest. Therefore, diversity has been a very important measure to evaluate the recommendation system \cite{Adomavicius2012Improving}. However, existing network embedding methods seldom consider the diversity. (2) Dynamic changes of user interest. User's preference is dynamic and how to model such a dynamic property is another challenge. (3) Scalability of recommendation system. Existing recommendation scenario often has a huge number of users and items, which is a serious problem with a scale beyond most of existing network embedding methods. 

To address these challenges, we propose \methodfull{}, a scalable \underline{R}ecommendation-based \underline{N}etwork \underline{E}mbedding method. In our method, when discovering the local structure of a user, we will not model all the items the user clicked. Instead, we propose a sampling method, which considers the diversity and dynamics of the user interest, to sample a portion of the items the user has clicked as the user's neighbors. In this way, the sampling method not only can incorporate the important properties of recommendation, i.e. the diversity and dynamics, to improve the recommendation accuracy, but also reduce the computational complexity of the algorithm. Furthermore, we deploy the algorithm on a recommendation system based on the Parameter Server to do distributed and parallel computing, which further facilitates the large-scale training available. 

In summary, the contributions of the paper can be listed as follows:
\begin{itemize}
\item We propose a network-embedding-based recommendation method, named \methodfull{}. When modeling the local structures between the users and items, our method is able to incorporate the dynamics and diversity of the user interest to produce more accurate and diverse recommendation results.
\item We implement our recommendation algorithm in a distributed way based on parameter server, which jointly makes the system available for billion-scale recommendation.
\item Experimentally, we deploy the whole system on a recommendation scenario of Taobao. Online A/B tests  demonstrate that our method is able to achieve more accurate results compared with CF and greatly improve the diversity of the recommendation results. Experiments on offline dataset Pinterest also demonstrate the quality of our method. 
\end{itemize}

\vspace{-25pt}
\begin{table*}[htb]
\centering\caption{ Multifaceted Comparisons between different methods }
\vspace{-5pt}
\label{tab:compare}
\begin{tabular}{|c|c|c|c|c|}
\hline
Method & Local-structure Preserving & Diversity  & Billion-scale & Complexity \\
\hline
GMF-CF/MLP-CF/NCF & $\times$ &  $\times$ &  $\times$    &  $O(|E|)$ \\
\hline
LINE/node2vec & ${\surd}$ &  $\times$  &  $\times$  & $O(|E|)$ \\
\hline
\methodfull{} & ${\surd}$ & ${\surd}$ & ${\surd}$ & $O(|V|)$ \\
\hline
\end{tabular}
\end{table*}
\vspace{-30pt}
\section{Related Work}
\vspace{-5pt}
\subsection{Collaborative Filtering}
Recommendation algorithms and systems are well-investigated research fields. In our work, we are only given the user-item interaction data. Therefore, we mainly introduce the CF-based recommendation methods and omit the discussions of content-based recommendation methods and the hybrid recommendation methods.

Collaborative Filtering exploits the interaction graph between the users and items to give the recommendation lists to users. Its basic idea is to match the users which have similar item preferences. Earlier CF methods mainly use the matrix factorization on the user-item matrices to obtain the latent user factors and item factors \cite{Deshpande2004Item,Harvey2011Bayesian,Sarwar2001Item}. The user factors and item factors together aim to reconstruct the original user-item matrices. However, the matrix factorization is just the linear-based methods, which is difficult to capture the user-item relationships. To overcome such a drawback, following works use the deep neural networks to perform collaborative filtering \cite{Wu2016Collaborative,Strub2015Collaborative}. However, most of the CF-based methods only aim to model the pairwise relationships between the user and item but omit their local structures. And many graph-based works have demonstrate that local structures like second-order relationships are very important for capturing graph structures \cite{tang2015line}. In this way, existing CF-based methods are sub-optimal for capturing the relationships between user and items.
\vspace{-15pt}
\subsection{Network Representation Learning}
Network embedding has been demonstrated as an effective methods for modeling local and global structures of a graph. It aims to learn a low-dimensional vector-representation for each node. DeepWalk \cite{perozzi2014deepwalk} and Node2vec \cite{Grover2016node2vec} propose to use the random walk and skip-gram to learn the node representations. LINE \cite{tang2015line} and SDNE \cite{Wang2016Structural} propose explicit objective functions for preserving first- and second-order proximity. Some further works \cite{Ou2016Asymmetric,Cao2015GraRep} use the matrix factorization to factorize high-order relation matrix. Aforementioned methods are designed for homogeneous networks. Then some following embedding methods for heterogeneous networks are proposed, like Metapath2vec \cite{dong2017metapath2vec} , HNE \cite{chang2015heterogeneous}, BiNE \cite{gao2018bine} and EOE \cite{xu2017embedding}. Some works further focuses on knowledge graph embedding \cite{xiao2016transg}. Although these network embedding methods are able to preserve the local structures of  the vertices, most of them are not specifically designed for the task of recommendation. They do not consider some specific properties of the recommendation tasks like the diversity and dynamic changes of user interest, the scalability issues of large-scale recommendation tasks. Therefore, how to propose an effective network embedding method for billion-scale recommendation is still an open problem.

In summary, we compare our method and the related works in Table \ref{tab:compare}. Our method is specifically designed for the recommendation scenario and thus consider some specific properties. Furthermore, the proposed method is very scalable and thus can apply to billion-scale recommendations. 
\vspace{-15pt}
\section{The Methodology}
\vspace{-10pt}
In our scenario, we have a large number of users and items. Each user may have different ways to interact with the items. For example, the user may view the items, collect the items or buy the items. In this way, we can build the user-item interaction graph, formally formulated as $G=(\mathcal{U},\mathcal{I},E)$. Here $\mathcal{U}$ denotes the total of $n$ users and $\mathcal{I}$ denotes the total of $T$ items. $\mathcal{U} \cup \mathcal{I}$ denotes the set of nodes in $G$. If a user $u\in \mathcal{U}$ views, collects or buys an item $i \in \mathcal{I}$, there is an edge $E_{ui}$ between $u$ and $i$. We use $E(v), v \in \mathcal{U} \cup \mathcal{I}$ to denote the edges connected to the node $v$. We assume that $G$ is connected. The recommendation problem is that given a user $u$, we hope to recommend some personalized items to the user based on his previous behavior. 
\vspace{-15pt}
\begin{figure}[htb]
\centering
\includegraphics[width=0.6\textwidth,height=0.2\textheight]{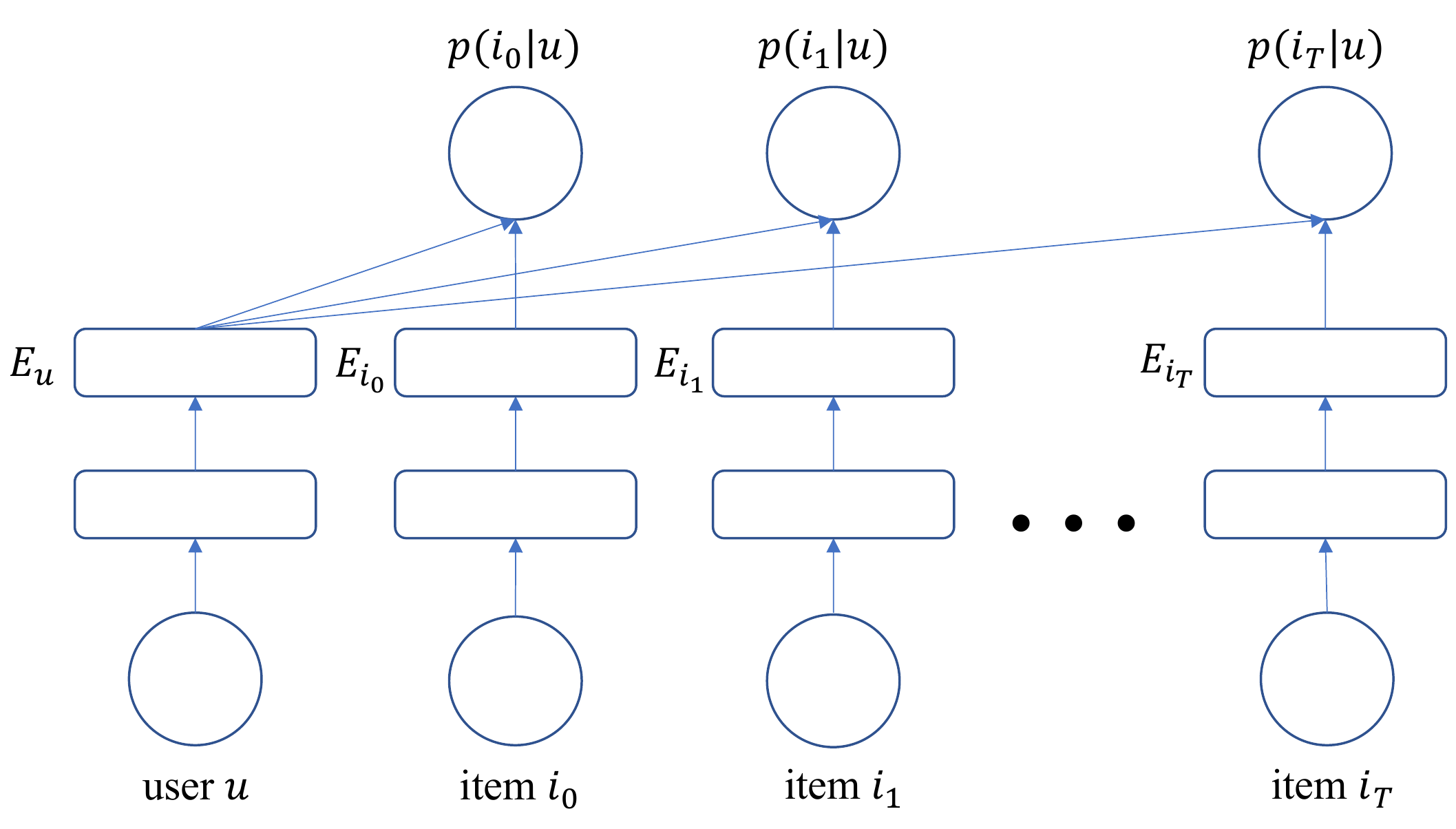}
\vspace{-10pt}
\caption{The framework of \methodfull{}.}
\label{pic:model_framework}
\vspace{-20pt}
\end{figure}
\vspace{-20pt}
\subsection{Network Embedding for Recommendation}
\vspace{-10pt}
Given the user-item interaction graph $G=(\mathcal{U},\mathcal{I},E)$, we aim to map each user and item to a common low-dimensional latent space, where user $u$ can be embedded as $\mathbf{E}_\mathcal{U}^u \in R^d$ and item $i$ can be embedded as $\mathbf{E}_{\mathcal{I}}^i \in R^d$. Then with the embeddings for each user and item, we can retrieve the similar items for the user as his recommendation results. 

To achieve this, we propose our method, whose framework can be shown in Figure \ref{pic:model_framework}. It consists of the embedding-lookup layer, embedding layer and softmax layer. The embedding-lookup layer helps us obtain the embeddings for the users and items. The embedding layer and softmax layer together model the interactions between the users and items to update the embedding-lookup layer. Then we introduce the designed loss functions to update the embeddings.  

We first consider how to model the local structure of a user in the given user-item graph. In the original space, the empirical distributions given a user can be defined as:
\begin{equation}
\hat{p}(i|u)=\frac{w_{ui}}{d_u}, \nonumber
\label{eqn:real_cond_prob}
\vspace{-5pt}
\end{equation}
where $w_{ui}$ is the weight between user $u$ and item $i$ and $d_u$ is the degree of user $u$. 

Then we hope to estimate the local structure of a user in the embedding space. Word2vec \cite{mikolov2013efficient} inspires us to use the inner product between two vertices to model their interactions. Then in our work, given a user $u$, we define the probability of item $i$ generated by user $u$ as: 
\begin{equation}
p(i|u)=\frac{exp({\mathbf{E}_{\mathcal{U}}^{u^T} \mathbf{E}_{
\mathcal{I}}^i})}{\sum_{j=1}^{|I|}{exp({\mathbf{E}_{\mathcal{U}}^u}^T \mathbf{E}_{\mathcal{I}}^j)}},
\label{eqn:cond_prob}
\end{equation}
where $T$ means the transpose of a matrix. 

Eqn. \ref{eqn:cond_prob} is a softmax-like loss function, which defines the conditional distributions $p(\cdot|u)$ of user $u$ over its neighborhoods, i.e. the entire item set, in the embedding space. 

With the empirical distributions on the original network and reconstructed distributions on the embedding space, we can learn the embedding by making the defined probability $p(\cdot|u)$ specified by the low-dimensional representations be close to the empirical distributions $\hat{p}(\cdot|u)$. We use the KL-divergence to measure the distance between the distributions. Then the loss functions can be defined as:
\begin{eqnarray}
L=\sum_{u\in\mathcal{U}}{\lambda_uKL(\hat{p}(\cdot|u),p(\cdot|u))} \propto -\sum_{(u,i)\in E}{w_{ui}logp(i|u)}, 
\label{eqn:ori_loss}
\vspace{-10pt}
\end{eqnarray}
where $\lambda_u$ denotes the prestige of user $u$ and we set $\lambda_u=d_u$.

Minimizing Eq. \ref{eqn:ori_loss} will make the vertices with similar neighbors similar to each other. Therefore, it can not only model the observed links on the graph, but also preserve the local structures for each node. 

\vspace{-15pt}
\subsection{Recommendation-based Sub-sampling}
\vspace{-5pt}
However, aforementioned network embedding meets two challenges for large-scale recommendation: (1) Minimizing Eqn. \ref{eqn:ori_loss} is time-consuming since for each edge it needs to run over the entire set of the items when evaluating $p(i|u)$. In this way, the whole complexity is $O(|E||\mathcal{I}|)$, which is unbearable for real recommendation systems. (2) Minimizing Eqn. \ref{eqn:ori_loss} only considers the topology of the graph. It does not consider the diversity and the time decay of the user interest, which are very important properties for recommendation systems. 

To reduce the complexity, we first adopt negative sampling as many methods do \cite{tang2015line}. For each positive edge $(u,i)$, we will sample some negative edges according to predefined distributions $P_{ui}$. By performing negative sampling, the objective function for each edge $(u,i)$ can be reformulated as: 
\vspace{-10pt}
\begin{equation}
L_{ui} = log(\sigma({\mathbf{E}_{\mathcal{U}}^u}^T \mathbf{E}_{\mathcal{I}}^i))+\sum_{j=1}^k E_{i_j \sim P_{u}}{(log(\sigma({\mathbf{E}_{\mathcal{U}}^u}^T E_{\mathcal{I}}^{i_j})))},
\label{eqn:final_loss}
\vspace{-10pt}
\end{equation}
where $k$ is the number of negative samples for each user-item pair, $P_{ui} \propto d^{3/4}_i$.

Although negative sampling can reduce the time complexity from $O(|E||\mathcal{I}|)$ to $O(k|E|)$, for billion-scale recommendation, a complexity linear to the number of edges is still a great challenge. 

To further reduce the complexity, we only select a portion of the items the user has clicked to obtain his behavior sequence. Then the question comes to how to select the items to effectively represent the user's interest. Here, we mainly consider two properties specified for recommendation. (1) The diversity of user interest: User interest is always diverse. A user will always focus on the items of more than one cluster. (2) The time decay of user interest: User interest is always dynamic. More recent user behavior is more reliable to reflect the recent user interest. Therefore, we should more focus on recent user behavior. Based on these two considerations, we define the selection probability for each user-item pair $(u,i)$ as follows:
\begin{equation}
p(u,i) = 0.999^{t_i}*click(u,c_i)^\gamma,
\label{eqn:prob_pos}
\end{equation}
where $t_i$ is the hours of the item $i$ from the most recent item, $c_i$ is the cluster index of item $i$ and $click(u,c_i) = \sum_{j \in c_i}{w_{uj}}$, $\gamma$ is set to $-0.2$. Then for each user, we will sample $m$ samples according to the defined probability in Eqn. \ref{eqn:prob_pos} to represent his behavior sequence. Then in this way, the complexity can be reduced from $O(k|E|)$ to $O(km|\mathcal{U}|)$, which is linear to the number of nodes. 

In summary,  on the one hand, if a user more recently shows the interest to an item, the item should have a larger probability to be sampled. On the other hand, the method is prone to sample the items of the clusters clicked less times by the user. In this way, our method may cover more clusters to ensure the diversity. Therefore, such a sampling strategy can simultaneously model the diversity and time decay of the user interest Furthermore, with the sampling strategy, we do not need to model all the edges in one iteration but instead for each user we only model a portion of its preferred items as the user's behavior sequence. It significantly reduce the time complexity. 

\vspace{-15pt}
\subsection{Implementation}
\vspace{-5pt}
In this section, we will introduce the technical implementation of the proposed \methodfull{}. The whole process can be divided into two phases: offline model training and online retrieval. The proposed \methodfull{} is implemented and deployed on KunPeng platform \cite{zhou2017kunpeng}. This section will describe them in detail.
\vspace{-10pt}
\subsubsection{Off-line Model Training}
To train the proposed \methodfull{}, we utilize the Stochastic Gradient Descent (SGD) on the loss function of Eqn. \ref{eqn:final_loss} to update the node embeddings. In detail, we use $E_{pos}$ to denote all the positive edges sampled by the method we proposed before. Then for each $(u,i)\in E_{pos}$, we can update their embeddings as follows: 
\begin{equation}
\mathbf{E}_{\mathcal{U}}^u = \mathbf{E}_{\mathcal{U}}^u+\lambda \{\sum_{z\in\{i\}\cup N_{neg}^k(u)}{[I(z,u)-\sigma{({\mathbf{E}_{\mathcal{U}}^u}^T \mathbf{E}_{\mathcal{U}}^z)}]\cdot \mathbf{E}_{\mathcal{U}}^z}\}, \\
\vspace{-5pt}
\label{eqn:update}
\end{equation}
where $I(a,b)$ is the indicator function that if $a=b$, $I(a,b)=1$, otherwise $I(a,b)=0$. $N_{neg}^k(i)$ is the negative neighborhoods of vertex $i$. $\lambda$ denotes the learning rate. Similarly, we can update embedding $E_{\mathcal{I}}^i$ for an item $i$ in a similar way, which we will not discussed more. 

From Eqn. \ref{eqn:update}, when given a positive edge, we can update their embeddings. Then we will go over all the pair of positive edges for several iterations to update their embeddings. The whole algorithm can be summarized in Alg. \ref{alg:train}.

\begin{algorithm}[htb]
\caption{Training Algorithm for \methodfull{}}
\label{alg:train}
\begin{algorithmic}[1]
\REQUIRE $G=(\mathcal{U},\mathcal{I},E)$
\ENSURE $\mathbf{E}_u$, $\mathbf{E}_I$
\STATE Initializing $E_u$ and $E_I$. 
% while-loop 
\WHILE{not converged} 
\STATE Construct the positive edge set $S_{pos}$ according to $G=(\mathcal{U},\mathcal{I},E)$ and Eq. \ref{eqn:prob_pos}.
\FORALL{$(u,i) \in S_{pos}$}
\STATE Construct the negative set $N_{neg}^k(u)$.
\STATE Update $\mathbf{E}_u$ and $\mathbf{E}_i$ according to Eq. \ref{eqn:update}.
\ENDFOR
\ENDWHILE 
\end{algorithmic}
\end{algorithm}
From Alg. \ref{alg:train}, we find that the learning process from Line $4$ to Line $6$ is independent for different edges, which inspires us to use some parallelization mechanism to implement it. Then we deploy the whole algorithm on the parameter server, which implements a data-parallelization mechanism. In detail, from Eqn. \ref{eqn:update} we find that to update a node's embedding, we only need to know the node's previous embeddings, the node's neighborhoods and their embeddings. Therefore, we can resort to parameter server to implement such a process in a parallelized way. The main workflow of the system is built as follows: (1) In each iteration, the server will assign each worker a subset of the vertices of the graph $G$. (2) Each worker will pull the assigned vertices from the server and calculate the positive and negative neighborhoods for the assigned vertices. Then with positive and negative sets, each worker can update the embeddings of the assigned vertices according to Eqn \ref{eqn:update}. (3) After updating, each worker will push his assigned vertices' embddings to the server. Such a training process will be iterated several times. 

\vspace{-15pt}
\subsubsection{Online Efficient Nearest Neighbor Search}
For online recommendation, we use the nearest neighbor search on the learned embedding space to make recommendations. That is, given a query user $u$, we can recommend items whose embeddings are the $K$-nearest-neighbors ($K$-nn) of the query user's embedding $E_u$. To achieve the $K$-nn search, we use the Faiss library \cite{johnson2017billion}， which is an efficient implementation for state-of-the-art product-quantization methods. Given that \methodfull{} is trained offline and all the user and item embeddings are computed via Parameter Server and saved in database, the efficient $K$-nn search enables the system to recommend items online. 
\vspace{-15pt}
\section{Experiments}
\vspace{-10pt}
The goal of \methodfull{} is to produce high-quality and scalable recommendations for real-world systems. Therefore, we conduct comprehensive experiments in two ways: Online A/B tests and Offline experiments. 
\vspace{-15pt}
\subsection{Datasets}
\vspace{-5pt}
We use two real-world datasets, i.e. Ali-mobile taobao and Pinterest in this paper.
\vspace{-10pt}
\begin{itemize}
\item Ali-mobile taobao: It is a mobile recommendation scenario deployed on Taobao, the largest E-commerce platform in China. The dataset is extremely large. It has about $1$ billion users, tens of million items and a total of about one hundred billion edges. Each edge denotes whether the user has clicked the products. We deploy our algorithm on the service to do online A/B test to evaluate our method.
\item Pinterest: The dataset is an image recommendation dataset constructed by \cite{geng2015learning}. We filter the users which have very few interactions with the items and only retain the users which have more than $20$ interactions. After the pre-processing, the dataset consists of $50$ thousand users, $10$ thousand items and $1.5$ million user-item edges. Each edge denotes whether the user has pinned the items. 
\end{itemize}
\vspace{-10pt}
\subsection{Online A/B tests}
The ultimate goal of the recommendation system is to lift the user's interest in the items. Therefore, we perform random A/B experiments on Ali-mobile taobao to demonstrate this, where a random set of users obtain the recommendation results of \methodfull{} and another obtain the results of CF-based methods. Any difference in the engagement of the items between the two groups can truly reflect the recommendation quality of two methods. Note that here we only use one baseline because deploying many methods online to do A/B tests will cost a lot of resources. And the reason why we choose CF is that it is well investigated for recommendation and existing network embedding methods cannot scale to billion-scale dataset. For more comparisons with state-of-the-art methods, we do offline experiment, which we will introduce in detail later. 

We use the following six metrics to measure the recommendation quality. 
\begin{itemize}
\item AVD (Averaged View Depth): The metric denotes how deep a user views the page. It measures the recommendation quality.  
\item ACN (Averaged Click Number): The metric measures the number of clicks on the items for each user in average. It measures the recommendation quality. 
\item P-CTR (Page Click-through Rate): for a page $p$, $pctr=\frac{\#click-throughs(p)}{\#impressions(p)}\times 100\%$. It measures the recommendation quality. 
\item U-CTR (User Click-through Rate): for a user $u$, $uctr=\frac{\#click-throughs(u)}{\#impressions(u)}\times 100\%$. It measures the recommendation quality. 
\item Re-C (Recommended number of clusters): The averaged number of clusters recommended to users. The clusters are obtained by using our clustering algorithm. The metric measures the recommendation diversity. 
\item CK-C (Clicked number of clusters): The averaged number of clusters clicked by users. It measures both the recommendation quality and the diversity. 
\end{itemize}

\vspace{-10pt}
\begin{table*}[htb]
\centering\caption{ Performance of online A/B tests on Ali-mobile taobao}
\vspace{3pt}
\label{exp:online}
\begin{tabular}{|c|c|c|c|c|c|c|c|c|}
\hline
Metrics & AVD & ACN & P-CTR & U-CTR & Re-C &  CK-C \\
\hline
Ali-mobile taobao & $9.54\%$ & $13.21\%$ & $4.99\%$ & $1.12\%$ & $20.49\%$ & $16.32\%$ \\
\hline
\end{tabular}
\vspace{-10pt}
\end{table*}

Table \ref{exp:online} summarizes the lift in engagement of items recommended by \methodfull{} compared with CF-based methods in controlled A/B experiments. From Table \ref{exp:online}, we have the following observations and analysis:
\begin{itemize}
\item We find that \methodfull{} can achieve a significant improvement in terms of AVD and ACN over the CF. It indicates by using the results of \methodfull{}, users are more willing to go deeper to view more items and click more items, which indirectly demonstrates the ranking quality of \methodfull{}.
\item In terms of the two CTR metrics, a popular and well-accepted metric to evaluate the recommendation quality, our proposed method also achieves a better result than CF. It further demonstrates that \methodfull{} is able to produce personalized items for users. The reason for a better recommendation quality is twofold. (1) Our method is able to capture the local structures between the users and items. (2) Our method considers the dynamic change of the user's interest.
\item We find that \methodfull{} achieves a higher Re-C compared with CF, which indicates that our recommended results are from more clusters. The reason is that our proposed method incorporates the diversity issue into the model design. 
\item More importantly, our method achieves a higher CK-C than CF, which demonstrates that not only our method can produce more diverse recommendations, but also the users are willing to click these diverse recommended items. It indicates that our method is able to improve the recommendation quality while improving the recommendation diversity. 
\item Under the billion-scale scenario, \methodfull{} can be deployed online and still obtain good results, which demonstrates the superiority of our method. 
\end{itemize}

%\begin{figure}[htb]
%\centering
%\includegraphics[width=0.48\textwidth]{pics/figure3.jpg}
%\caption{The ratio of the impressions in the top $20$ bucket accounting for the total impressions. The blue curve is the online result of previous method and the red curve is the online result of the proposed method \methodfull{}.}
%\label{pic:ratio_impressions}
%\end{figure}

%Furthermore, we conduct another experiment to see more properties of \methodfull{}. We sort the number of impressions for each item in descent order and divide them into $10000$ buckets. In each bucket, we observe the ratio of the impressions in the bucket accounting for the total impressions. We plot the impression ratio for the top $20$ largest buckets. The result is shown in Figure \ref{pic:ratio_impressions}. 

%From Figure \ref{pic:ratio_impressions}, we find that starting from the third bucket, the red curve is above the blue curve. It indicates that the impression flow moves from the top part to the middle part. It demonstrates that our proposed method can avoid recommending very popular items but produce diverse recommendation results. 
\vspace{-15pt}
\subsection{Showcase}
\vspace{-15pt}
\begin{figure*}[htb]
\centering
\subfigure[]{
\includegraphics[width=0.42\textwidth]{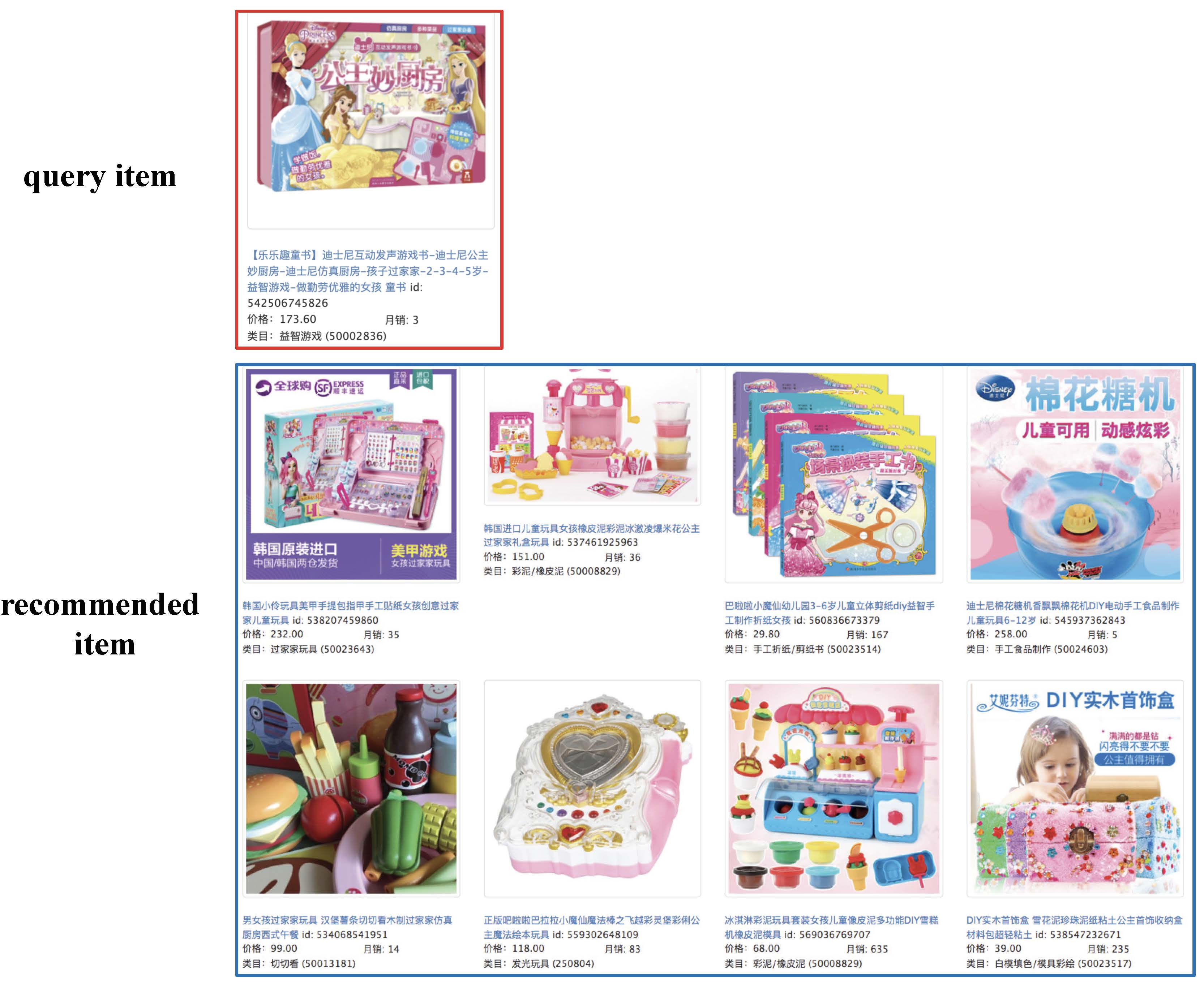}
\label{pic:showcase_1}}
\hspace{20pt}
\subfigure[]{
\includegraphics[width=0.4\textwidth]{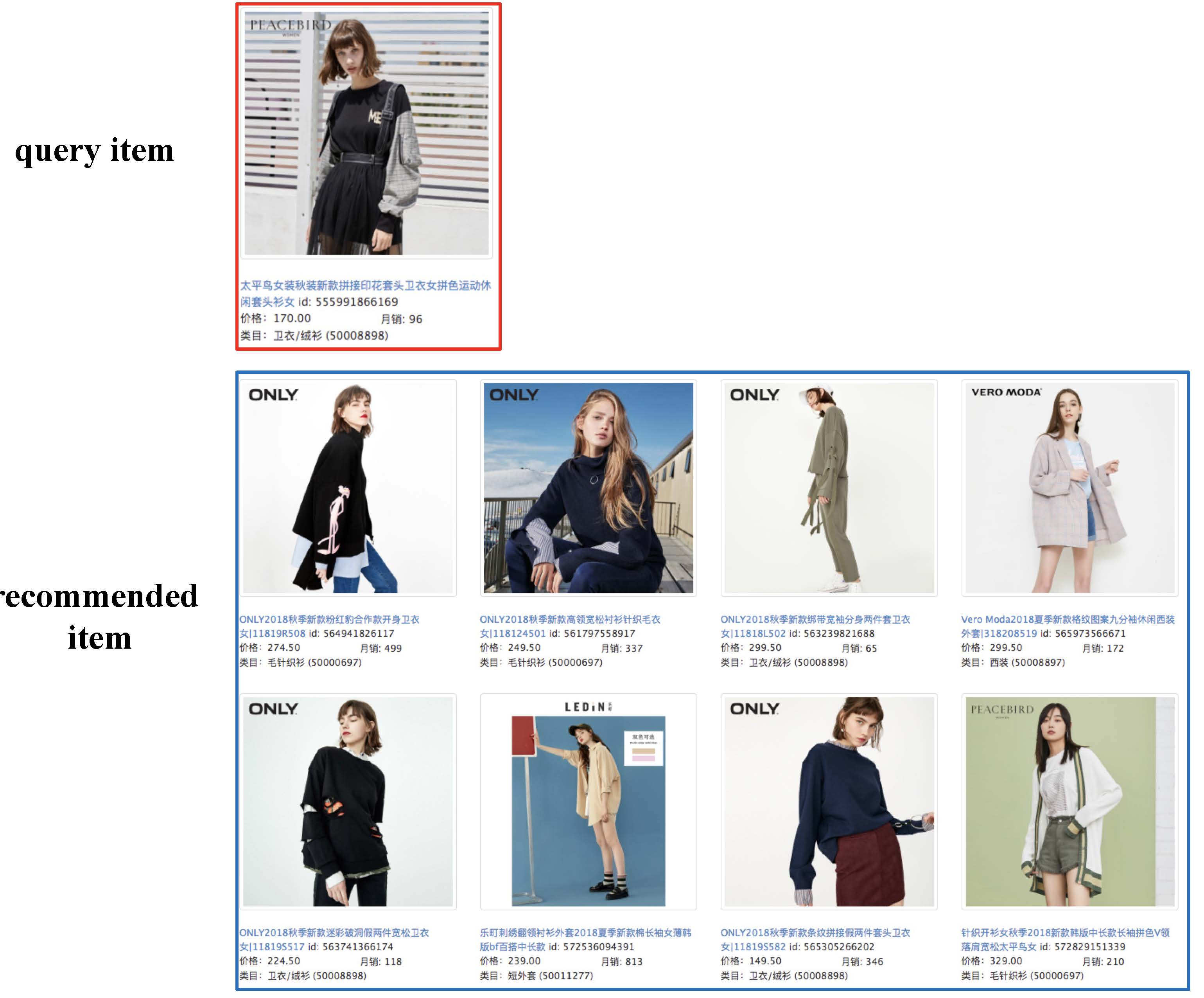}
\label{pic:showcase_2}}
\caption{ Real showcase on Ali-mobile taobao: Given an item (in red box), searching for the nearest $8$ items (in blue box) using the embeddings learned by \methodfull{}. }
\label{pic:showcase}
\vspace{-15pt}
\end{figure*}

In this section, we give some showcase to see some intuitions regarding the embeddings we learn. After the learning process of \methodfull{}, all the items will have embeddings. Then in this experiment, given a query item's embedding, we aim to find the most similar $8$ items whose embeddings have the smallest distance with the query. Then we display both the query image and the recommended images in Figure \ref{pic:showcase}.

In Figure \ref{pic:showcase_1}, the query item is a princess-style educational toy for girls. When we look at the returned results, these images belong to different categories with the query, like plasticine and origami. But all of them are for fun and a majority of them are also princess-style. It demonstrates that our method is able to find more categories of items but retain the primary style of the item. In Figure \ref{pic:showcase_2}, the query item is a woman sweatshirt of the brand of Peacebird. The returned images are all coat, sweater or sweatshirt of the brand of Peacebird or Only. Similarly, the returned images and the query image are all casual style but belong to different fine-grained categories. Moreover, actually the brand of Peacebird and Only have very similar styles and our method can learn their inherit relationships. 

In summary, in our method, we do not have the item features and the direct relationships between items. We are only given the user-item interaction graph. Although in this case, our method still can model the item relationships by using the user behaviors  as the bridge. It demonstrates that by using the network embedding method we propose, the learned embeddings can capture the local  relationships between the entities.

\vspace{-10pt}
\subsection{Offline Experiment}
\vspace{-15pt}

\begin{table*}[htb]
\centering\caption{ Recommendation Performance on Pinterest. }
\vspace{2pt}
\label{tab:pinterest}
\begin{tabular}{|c|c|c|c|c||c|c|c|c||c|c|c|c|}
\hline
\multirow{2}{*}{Method} &
 \multicolumn{4}{c||}{HR} &  \multicolumn{4}{c||}{NDCG} &  \multicolumn{4}{c|}{RR}  \\
\cline{2-13} & Top$5$ & Top$10$ & Top$50$ & Top$100$ & Top$5$ & Top$10$ & Top$50$ & Top$100$ & Top$5$  & Top$10$ & Top$50$ & Top$100$ \\
\hline
GMF & 0.501 & 0.678 & 0.9 & 0.97 & 0.332 & 0.386  & 0.425 &  0.434 & 0.276 & 0.292 &  0.301 & 0.307 \\
\hline
MLP & 0.504 & 0.679 & 0.908 & \textbf{0.99} & 0.341 & 0.385 & 0.426 & 0.436 & 0.275 & 0.295 & 0.302 & 0.309 \\
\hline
NCF & 0.529 & 0.688 & 0.912 & \textbf{0.99} & 0.35 & 0.405 & 0.443 & 0.454 & 0.298 & 0.32 & 0.343 & 0.341 \\
\hline
LINE & \textbf{0.536} & \textbf{0.7} & 0.91 & \textbf{0.99} & 0.353 & 0.409 & 0.448 & 0.455 & 0.297 & 0.325 & 0.334 & 0.335 \\
\hline
node2vec & 0.527 & 0.691 & \textbf{0.93} & \textbf{0.99} & 0.355 & \textbf{0.411} & \textbf{0.459} & \textbf{0.46} & \textbf{0.302} & \textbf{0.329} & 0.341 & 0.342 \\
\hline
\methodfull{} & 0.531 & 0.695 & 0.925 & \textbf{0.99} & \textbf{0.356} & 0.41 & 0.45 & 0.457 & 0.3 & 0.327 & \textbf{0.345} & \textbf{0.348} \\
\hline
\end{tabular}
\end{table*}

To compare more baselines to get comprehensive results, we conduct the offline experiments on the dataset of Pinterest. We randomly sample 90$\%$ user-item pairs as the training set and the rest as the testset. For training set, we use $9$-fold cross-validation to tune the parameters for all the methods. Note that in this dataset, we do not have the cluster and time information for the item. So we uniformly sample the items to do training. To evaluate the performance, we use the following three metrics: Normalized Discounted Cumulative Gain (NDCG), Mean Reciprocal Rank (MRR) and Hit Rate (HR). NDCG and MRR will consider the rank of the hit and will assign higher scores to hits at top ranks. While HR will only evaluate whether the test items are hit or not. We calculate all the metrics for the test users and report the average score. 

We first use the advanced CF-based methods GMF, MLP and NCF \cite{He2017Neural} as baseline methods. We perform the same process of parameter search as the work \cite{He2017Neural} did to select the optimal parameters. For network embedding methods, since we only have the graph topology, in this case LINE \cite{tang2015line} and node2vec \cite{Grover2016node2vec} are state-of-the-art network embedding methods, so we choose them as the baselines. For LINE, we use LINE$_{1st+2nd}$ with the default parameter settings. For node2vec, we also use the default settings except for the bias parameters $p,q$, which we conduct the grid search from $\{0.5,1\}$. The embedding dimension of them is all set as 128. 

The results are shown in Table \ref{tab:pinterest}. From Table. \ref{tab:pinterest}, we find that \methodfull{} achieves a better performance than all the CF-based methods. The reason is that \methodfull{} is able to capture the local structure of each user while CF-based methods only focus on the direct links the user has clicked. It demonstrates that capturing the local structures on the user-item graph is important for recommendation. LINE, node2vec and \methodfull{} achieve similar performance in different evaluation metrics and scenarios. But our method runs much faster than node2vec and LINE, which will be discussed later. Therefore, \methodfull{} is a better balance between accuracy and efficiency.

\begin{figure}[htb]
\centering
\includegraphics[width=0.5\textwidth]{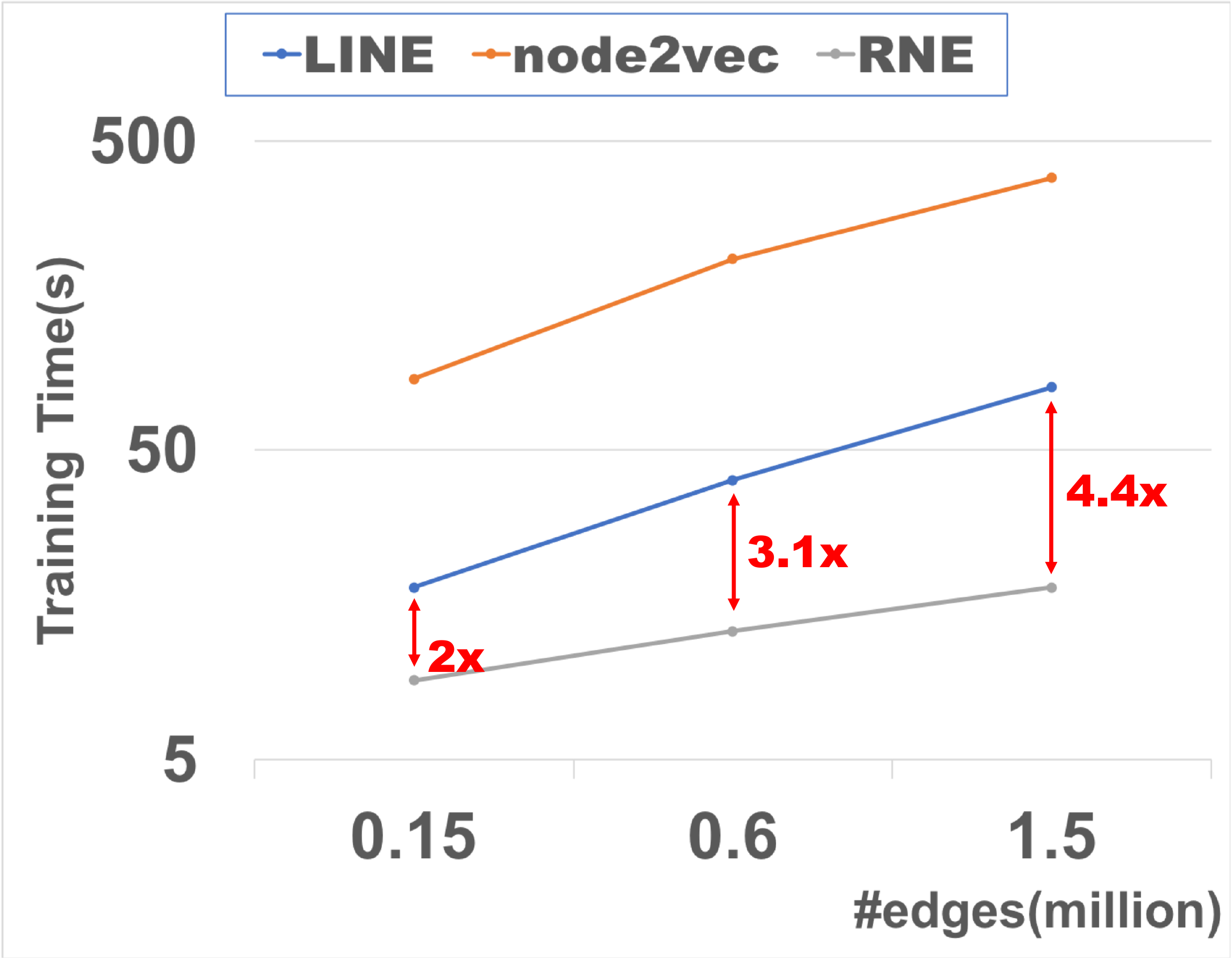}
\caption{ Time comparisons on Pinterest dataset. We change the number of edges to be trained and report the training time for each network embedding method. }
\label{pic:time}
\vspace{-15pt}
\end{figure}

Now we discuss the training time of LINE, node2vec and \methodfull{}. For a fair comparison, we do not use the distributed strategy for \methodfull{}. From Figure \ref{pic:time}, we find that \methodfull{} can boost the running time over LINE and node2vec. Specifically, when the training edges increase from $0.15$ million to $1.5$ million, the running time improvement of \methodfull{} compared with LINE will be larger and larger, from $2$x to $4.4$x. When the edges continuously increase to the billion-scale dataset like the Ali-mobile taobao dataset, it is difficult for LINE and node2vec to obtain the results. But \methodfull{} can still obtain a good result. The reasons why our method can scale to billion-scale dataset are twofold: (1) The proposed sampling method avoids us running over all the edges in the graph. (2) Our method can be deployed on distributed system for parallel computations. 

In summary, \methodfull{} has a good scalability, which is much more efficiency than baseline methods and can scale to billion-scale recommendation scenario, meanwhile \methodfull{} do not sacrifice its recommendation accuracy. 

\vspace{-15pt}
\section{Conclusion}
\vspace{-10pt}
In this paper, we propose a novel network embedding method named \methodfull{} for scalable recommendation. The proposed network embedding method is able to capture the local structures on the user-item graph to achieve a better recommendation quality. Specifically, to consider the specific properties for recommendation, i.e the diversity and time-decay of user interest, we design a sampling method for embedding process to incorporate these properties. And the sampling method also guarantees the scalability of the proposed method while almost preserving the recommendation quality. We also deploy our algorithm on parameter server to make it available for large-scale recommendation. Experimental results on online A/B tests and offline experiments all demonstrate the superiority of the proposed method. 

For the future work, we may consider the user and item features, which can further address the sparsity and cold-start problem. We also want to analyze the role of features and topology structures for recommendation. 
\vspace{-15pt}

\section{Acknowledgement}
We would like to thank all the colleagues of our team and all the members of our cooperative team: the search engine team in Alibaba. They provide many helpful comments for the paper. We also would like to thank the support of the Initiative Postdocs Supporting Program and the valuable comments provided by all the reviewers.

\bibliographystyle{splncs04}
\bibliography{sigproc}

\end{document}